\documentclass[sigconf]{acmart}

\AtBeginDocument{%
  \providecommand\BibTeX{{%
    \normalfont B\kern-0.5em{\scshape i\kern-0.25em b}\kern-0.8em\TeX}}}

\copyrightyear{2020} 
\acmYear{2020} 
\setcopyright{acmlicensed}\acmConference[UbiComp/ISWC '20 Adjunct]{Adjunct Proceedings of the 2020 ACM International Joint Conference on Pervasive and Ubiquitous Computing and Proceedings of the 2020 ACM International Symposium on Wearable Computers}{September 12--16, 2020}{Virtual Event, Mexico}
\acmBooktitle{Adjunct Proceedings of the 2020 ACM International Joint Conference on Pervasive and Ubiquitous Computing and Proceedings of the 2020 ACM International Symposium on Wearable Computers (UbiComp/ISWC '20 Adjunct), September 12--16, 2020, Virtual Event, Mexico}
\acmPrice{15.00}
\acmDOI{10.1145/3410530.3414374}
\acmISBN{978-1-4503-8076-8/20/09}

\begin{document}

\title{Alexa Depression and Anxiety Self-tests: A Preliminary Analysis of User Experience and Trust}

\author{Juan C. Quiroz}
\orcid{0000-0002-2831-7689}
\affiliation{%
  \institution{Centre for Big Data Research in Health, UNSW}
}
\affiliation{%
  \institution{Australian Institute of Health Innovation, Macquarie University}
  \city{Sydney}
  \state{Australia}
}
\email{juan.quiroz@unsw.edu.au}

\author{Tristan Bongolan}
\affiliation{%
  \institution{Macquarie University}
  \city{Sydney}
  \state{Australia}
}
\email{tbongolan@hotmail.com}
  
\author{Kiran Ijaz}
\orcid{0000-0001-8722-6595}
\affiliation{%
  \institution{Australian Institute of Health Innovation, Macquarie University}
  \city{Sydney}
  \state{Australia}
}
\email{kiran.ijaz@mq.edu.au}

\renewcommand{\shortauthors}{Quiroz et al.}

\begin{abstract}
Mental health resources available via websites and mobile apps provide support such as advice, journaling, and elements from cognitive behavioral therapy. 
The proliferation of spoken conversational agents, such as Alexa, Siri, and Google Home, has led to an increasing interest in developing mental health apps for these devices. 
We present the pilot study outcomes of an Alexa Skill that allows users to conduct depression and anxiety self-tests. 
Ten participants were given access to the Alexa Skill for two-weeks, followed by an online evaluation of the Skill's usability and trust.
Our preliminary evaluation suggests that participants trusted the Skill and scored the usability and user experience as average. 
Usage of the Skill was low, with most participants using the Skill only once.
In view of work-in-progress, we also present a discussion of implementation and study design challenges to guide the current literature on designing spoken conversational agents for mental health applications.

\end{abstract}



\maketitle

\section{Introduction}


Mental health problems are a growing global challenge affecting people of all different backgrounds, ages, and socioeconomic status~\cite{woodward_beyond_2019}. 
To tackle this challenge, mental health resources are increasingly available online and via mobile apps~\cite{kien_fully_2017,morris_efficacy_nodate,kamita_chatbot_2018}. 
The proliferation of conversational agents has made them attractive for health applications~\cite{laranjo} and mental health~\cite{vaidyam_chatbots_2019}.
Notable chatbots that monitor mood and use aspects of CBT to help users deal with anxiety and depression include Woebot~\cite{stiles-shields_woebot_2019} and TESS~\cite{fulmer_using_2018}.




The advancements and rapid adoption of spoken conversational agents, such as Siri and Alexa, make them attractive as a channel for providing mental health resources to users~\cite{maharjan_hear_2019}. 
Spoken conversational agents interact with users via spoken natural language. 
For mental health applications, this requires users to vocalize responses about their mental health status, which is different than typing responses to a chatbot or on a website.
One study explained that some people are more likely to have truthful interactions about their mental health with technology than wth mental health professionals~\cite{ravichander_proceedings_2018}.  

This paper presents work-in-progress findings of an Alexa Skill we developed that performs depression and anxiety self-tests.
Current Alexa Skills focus on guiding, educating, and helping users manage mental health issues.
Some Alexa Skills examples include management for anxiety and stress through advice sessions (\textit{Anti Anxiety}, \textit{Anxiety Stress}), assisting people with depression by providing tasks to boost their mood (\textit{Mental Health Day Manager}), management advice and education for children and teenagers dealing anger, stress, anxiety, and depression (\textit{Mental Health Spies}), and targeted exercises depending on the situation (work, studies, life) causing the user stress (\textit{Mindscape}).
One study used Alexa to monitor a user's mental health behaviors and symptoms, requiring users to self-report data on sleep, mood, and activity levels~\cite{maharjan_hear_2019}. 
However, further studies are required to provide evidence regarding the efficacy of delivering mental health resources via spoken conversational agents.

Our contributions are: (1) We developed an Alexa Skill that allows users to conduct depression and anxiety self-tests and makes exercise recommendations to alleviate anxiety and depression symptoms; (2) We conducted a pilot study with 10 participants to assess the usability of our Alexa Skill.

\section{Methodology}

\subsection{Study Design}
In this preliminary study, we recruited 10 participants who owned an Alexa device or had a smartphone where the Alexa app could be installed. 
Participants first completed an online questionnaire that collected demographics and conversational agent usage habits. 
The questionnaire also included depression (PHQ-9~\cite{kroenke_phq-9_2001}) and anxiety (GAD-7~\cite{jordan_psychometric_2017}) self-tests.
This helped familiarize the participants with the depression and anxiety self-tests that they would later complete with our Alexa Skill. 
In our results, we compare online self-test scores vs the self-test scores completed using our Alexa Skill.

Participants were given access to the Alexa Skill for two weeks.
It was recommended to participants to use the Skill regularly, with reminders sent every three days. 
After two weeks, we asked participants to complete another questionnaire, which included questions to assess usability, user experience, and trust.
Ethics approval was granted by Macquarie University’s Human Research Ethics Committee for Medical Sciences (ethics reference number of 52020662417083). 

\subsection{The Alexa Skill}
The Alexa Skill allowed users to express their emotions, conduct self-tests for depression and anxiety, and make a number of suggestions to improve the user's current state-of-mind. 
Each session began with Alexa asking the user how they were feeling. 
After expressing their emotions, the user was prompted to either complete self-tests for depression and anxiety or to hear self-help exercises.  

If the user expressed a symptom related to anxiousness or depression, Alexa prompted them to complete a self-test depending on their listed emotions. 
After the user completed the depression and anxiety self-tests, Alexa stated their depression and anxiety scores.
Afterward, Alexa would recommend that the user practice one of five actions (randomly selected): breathing exercise; muscle relaxation exercise; lifestyle recommendations such as proper sleep, exercise, and maintaining a healthy diet; journaling; practice gratitude.


\subsection{Evaluation}

The questionnaire after the two-week period collected the participant's final self-test scores for depression and anxiety, a perspective of the app’s usability, user experience, trust, and feedback. 
We used the following questionnaires to assess the Alexa Skill: the System Usability Scale (SUS) for system usability~\cite{borsci_dimensionality_2009}; the User Experience Questionnaire (UEQ) for pragmatic and hedonic qualities~\cite{hinderks_developing_2019}; and the Technology Trust Questionnaire for trust between the system and the user~\cite{jian_foundations_2000}.


\section{Results}

The average age of the participants was 21.6 years old and 8/10 were male.
9/10 of the participants used conversational agents. 
7/10 of the participants also indicated that they had not previously used an app for mental well-being.
After the two-week period, 7/10 participants indicated that they rarely used our Alexa Skill.

\begin{figure}[t]
  \centering
  \includegraphics[width=\linewidth,height=4.5cm]{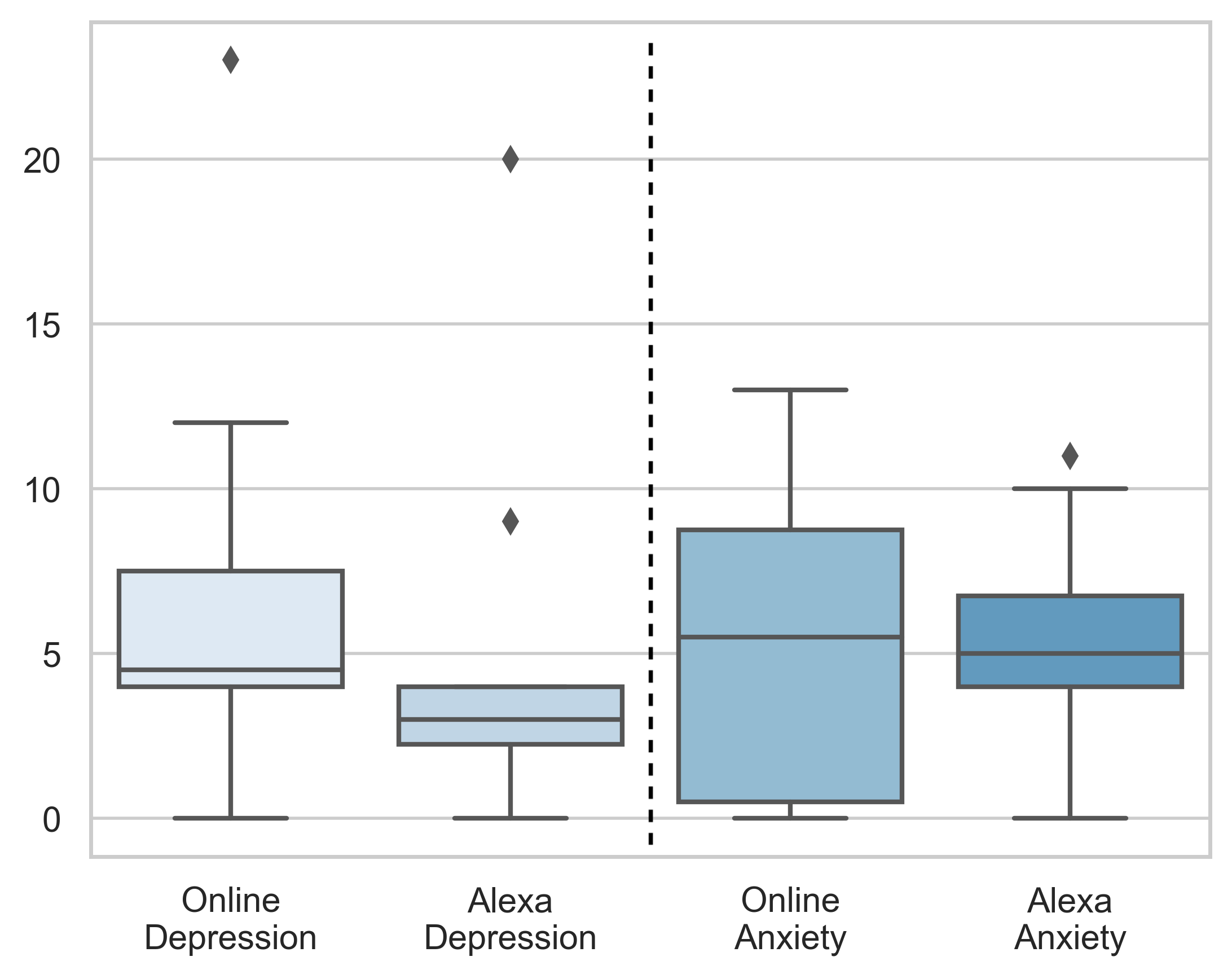}
  \caption{Depression (PHQ-9) and anxiety (GAD-7) self-test scores completed online (webpage) vs using the Alexa Skill.}
  \label{fig:gad}
\end{figure}

Figure~\ref{fig:gad} shows the anxiety and depression scores completed using a webpage (online) vs using Alexa. 
Given that the majority of the participants rarely used the Alexa Skill, we cannot attribute the change in anxiety score or depression scores to the use of the Alexa Skill or to the delivery of the self-tests via Alexa.

\begin{figure}[h]
  \includegraphics[width=\linewidth,height=3.5cm]{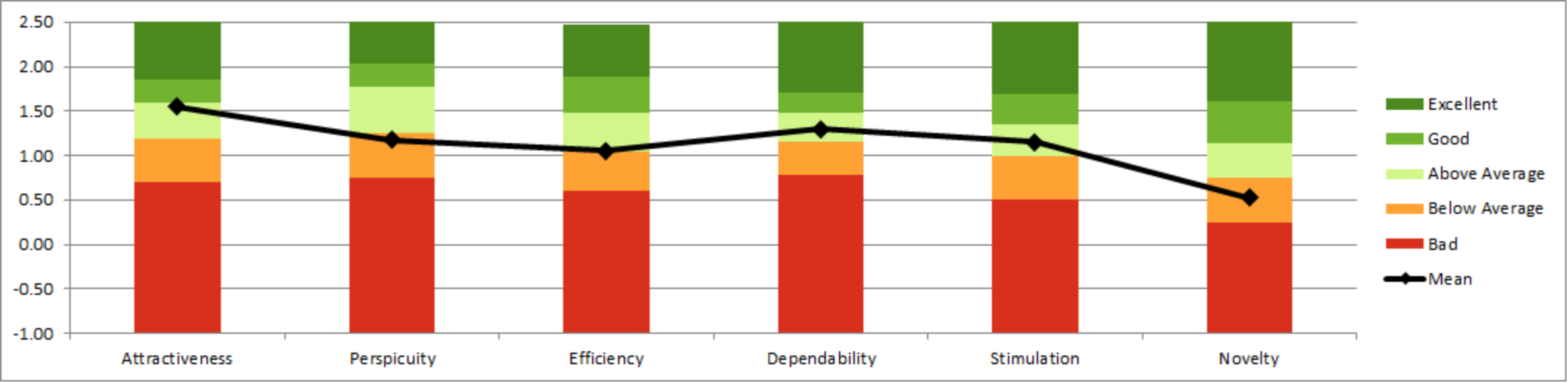}
  \caption{User experience scores of the Alexa Skill for depression and anxiety self-tests.} 
  \label{fig:ueq}
  \includegraphics[width=\linewidth,height=4.5cm]{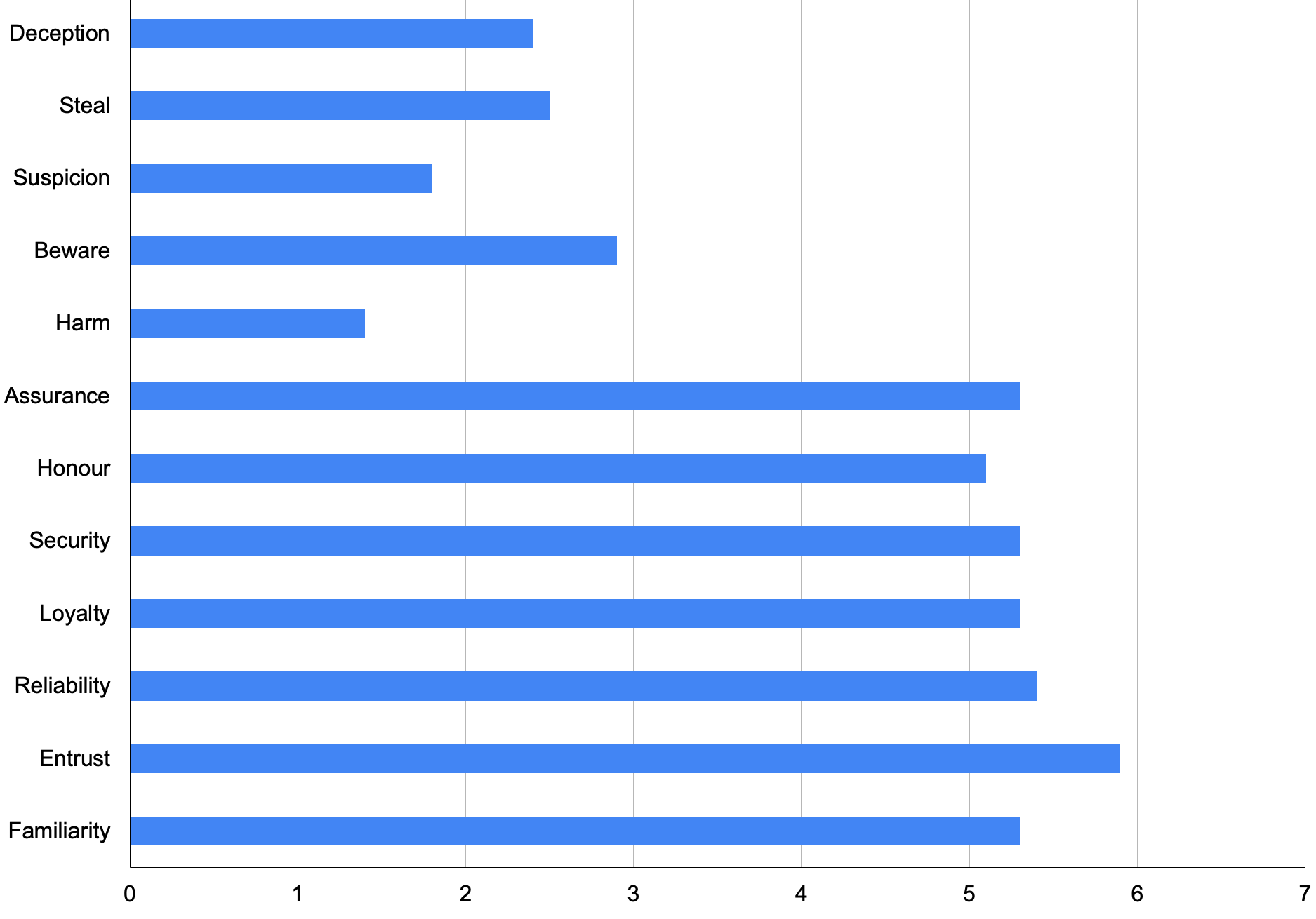}
  \caption{Trust scores of the Alexa Skill.}
  \label{fig:trust}
\end{figure}

Figure~\ref{fig:ueq} shows the user experience scores of the Alexa Skill. 
The participants found the app's attractiveness, dependability, and stimulation to be above average, but attractiveness, perspicuity, efficiency, and novelty were scored below average.
Figure~\ref{fig:trust} shows the trust scores for the Alexa Skill.
We conclude, participants mostly trusted the app. 

\section{Discussion and Challenges}
This pilot showed a willingness from participants to trust Alexa with personal information such as depression and anxiety scores. 
The user experience scores of the Alexa Skill also showed that participants considered it lacking in efficiency and novelty.

The design of this study, the implementation of the Alexa Skill, and conducting the pilot, highlighted a number of challenges when it comes to the development of an Alexa Skill for the mental health space. 

\textbf{Cold Start Problem:} The first user interactions with an Alexa Skills with multi-step or branching dialogues can be challenging for new users.
New users do not know the possibility of dialogue flows, the intent recognitions that have been programmed into the Alexa Skill, and the responses that are valid (how to respond to a question from the Alexa Skill). 
While Alexa Skills can be deployed with a user manual or by walking the user through a tutorial, complex skills may require regular use and various trials before the user is comfortable interacting with the Alexa Skill. 
In our pre-pilot tests, users struggled when interacting with our Alexa Skill because they gave responses that were not captured by the intent recognition rules we had programmed. 
Careful design must be planned to ensure intuitive interactions between the user and the Alexa Skill, as any difficulties are bound to discourage users from future use. 
This is especially important when designing spoken conversational agent apps for mental health, where users may be experiencing distress when they decide to use the Skill.

\textbf{Robust Dialogues:} Alexa Skills with deep dialogues need robust handling of user responses.
In our Alexa Skill, the depression and the anxiety self-tests involved Alexa reading multiple questions and waiting for a user response to each question.
During testing, some users were frustrated by having to repeat the depression or the anxiety self-test because the speech recognition of Alexa did not understand their response, or because the Skill closed due to an Alexa error or an error in our Skill. 
If a user is experiencing distress, this type of experience may worsen the user's mental state or fail to provide the support intended by the app/Skill.

\textbf{Handling expression of emotions:} Our Alexa Skill allowed users to state how they were feeling using single words. 
This aspect of our Alexa Skill was brittle, since expressing a word not included in our list of emotions meant that Alexa would ask the user to repeat themselves. 
Ideally, our long-term goal is to support all expressions of emotions, with Alexa being able to acknowledge what the user is experiencing. 
We believe this acknowledgment can help users in isolation and experiencing distress, but it poses technical challenges in natural language understanding. 
Some of the users that tested our Skill also expressed their emotions by referring to physical symptoms associated with an emotion they were feeling, i.e. "sweaty palms", "heart racing", which poses the additional challenge of knowing the association between a body response and an emotion. 

\textbf{Engagement with the self-test Alexa Skill:}  Engagement with health support technologies is concerning. 
Related literature in mobile health apps for mental health reports users' uptake and engagement challenges~\cite{ngUserEngagementMental2019}. 
In our pilot, we observed similar trends where participants rarely used the Alexa Skill. 
Enhancing the engagement of conversational agents through storytelling~\cite{battaglinoIncreasingEngagementConversational2015}, personalization~\cite{Baki2019}, and affect~\cite{Callejas2011} has been proposed in the past.
However, further research is needed to understand the design implication of such features for spoken conversational agents in the mental health context. 


\bibliographystyle{ACM-Reference-Format}
\bibliography{references}


\begin{thebibliography}{19}


\ifx \showCODEN    \undefined \def \showCODEN     #1{\unskip}     \fi
\ifx \showDOI      \undefined \def \showDOI       #1{#1}\fi
\ifx \showISBNx    \undefined \def \showISBNx     #1{\unskip}     \fi
\ifx \showISBNxiii \undefined \def \showISBNxiii  #1{\unskip}     \fi
\ifx \showISSN     \undefined \def \showISSN      #1{\unskip}     \fi
\ifx \showLCCN     \undefined \def \showLCCN      #1{\unskip}     \fi
\ifx \shownote     \undefined \def \shownote      #1{#1}          \fi
\ifx \showarticletitle \undefined \def \showarticletitle #1{#1}   \fi
\ifx \showURL      \undefined \def \showURL       {\relax}        \fi
\providecommand\bibfield[2]{#2}
\providecommand\bibinfo[2]{#2}
\providecommand\natexlab[1]{#1}
\providecommand\showeprint[2][]{arXiv:#2}

\bibitem[\protect\citeauthoryear{Battaglino and Bickmore}{Battaglino and
  Bickmore}{[n.d.]}]%
        {battaglinoIncreasingEngagementConversational2015}
\bibfield{author}{\bibinfo{person}{Cristina Battaglino} {and}
  \bibinfo{person}{Timothy~W. Bickmore}.} \bibinfo{year}{[n.d.]}\natexlab{}.
\newblock \showarticletitle{Increasing the engagement of conversational agents
  through co-constructed storytelling}. In
  \bibinfo{booktitle}{\emph{{{INT}}/{{SBG}}@{{AIIDE}}}} (2015).
\newblock


\bibitem[\protect\citeauthoryear{Borsci, Federici, and Lauriola}{Borsci
  et~al\mbox{.}}{2009}]%
        {borsci_dimensionality_2009}
\bibfield{author}{\bibinfo{person}{Simone Borsci}, \bibinfo{person}{Stefano
  Federici}, {and} \bibinfo{person}{Marco Lauriola}.}
  \bibinfo{year}{2009}\natexlab{}.
\newblock \showarticletitle{On the dimensionality of the {System} {Usability}
  {Scale}: a test of alternative measurement models}.
\newblock \bibinfo{journal}{\emph{Cognitive Processing}} \bibinfo{volume}{10},
  \bibinfo{number}{3} (\bibinfo{date}{Aug.} \bibinfo{year}{2009}),
  \bibinfo{pages}{193--197}.
\newblock
\showISSN{1612-4782, 1612-4790}
\urldef\tempurl%
\url{https://doi.org/10.1007/s10339-009-0268-9}
\showDOI{\tempurl}


\bibitem[\protect\citeauthoryear{Callejas, López-Cózar, Ábalos, and
  Griol}{Callejas et~al\mbox{.}}{2011}]%
        {Callejas2011}
\bibfield{author}{\bibinfo{person}{Zoraida Callejas}, \bibinfo{person}{Ramón
  López-Cózar}, \bibinfo{person}{Nieves Ábalos}, {and}
  \bibinfo{person}{David Griol}.} \bibinfo{year}{2011}\natexlab{}.
\newblock \showarticletitle{Affective conversational agents: the role of
  personality and emotion in spoken interactions}.
\newblock In \bibinfo{booktitle}{\emph{Conversational agents and natural
  language interaction: {Techniques} and effective practices}}.
  \bibinfo{publisher}{IGI Global}, \bibinfo{pages}{203--222}.
\newblock


\bibitem[\protect\citeauthoryear{Fulmer, Joerin, Gentile, Lakerink, and
  Rauws}{Fulmer et~al\mbox{.}}{2018}]%
        {fulmer_using_2018}
\bibfield{author}{\bibinfo{person}{Russell Fulmer}, \bibinfo{person}{Angela
  Joerin}, \bibinfo{person}{Breanna Gentile}, \bibinfo{person}{Lysanne
  Lakerink}, {and} \bibinfo{person}{Michiel Rauws}.}
  \bibinfo{year}{2018}\natexlab{}.
\newblock \showarticletitle{Using {Psychological} {Artificial} {Intelligence}
  ({Tess}) to {Relieve} {Symptoms} of {Depression} and {Anxiety}: {Randomized}
  {Controlled} {Trial}}.
\newblock \bibinfo{journal}{\emph{JMIR mental health}} \bibinfo{volume}{5},
  \bibinfo{number}{4} (\bibinfo{date}{Dec.} \bibinfo{year}{2018}),
  \bibinfo{pages}{e64}.
\newblock
\showISSN{2368-7959}
\urldef\tempurl%
\url{https://doi.org/10.2196/mental.9782}
\showDOI{\tempurl}


\bibitem[\protect\citeauthoryear{Hinderks, Schrepp, Domínguez~Mayo, Escalona,
  and Thomaschewski}{Hinderks et~al\mbox{.}}{2019}]%
        {hinderks_developing_2019}
\bibfield{author}{\bibinfo{person}{Andreas Hinderks}, \bibinfo{person}{Martin
  Schrepp}, \bibinfo{person}{Francisco~José Domínguez~Mayo},
  \bibinfo{person}{María~José Escalona}, {and} \bibinfo{person}{Jörg
  Thomaschewski}.} \bibinfo{year}{2019}\natexlab{}.
\newblock \showarticletitle{Developing a {UX} {KPI} based on the user
  experience questionnaire}.
\newblock \bibinfo{journal}{\emph{Computer Standards \& Interfaces}}
  \bibinfo{volume}{65} (\bibinfo{date}{July} \bibinfo{year}{2019}),
  \bibinfo{pages}{38--44}.
\newblock
\showISSN{0920-5489}
\urldef\tempurl%
\url{https://doi.org/10.1016/j.csi.2019.01.007}
\showDOI{\tempurl}


\bibitem[\protect\citeauthoryear{Jian, Bisantz, and Drury}{Jian
  et~al\mbox{.}}{2000}]%
        {jian_foundations_2000}
\bibfield{author}{\bibinfo{person}{Jiun-Yin Jian}, \bibinfo{person}{Ann~M.
  Bisantz}, {and} \bibinfo{person}{Colin~G. Drury}.}
  \bibinfo{year}{2000}\natexlab{}.
\newblock \showarticletitle{Foundations for an {Empirically} {Determined}
  {Scale} of {Trust} in {Automated} {Systems}}.
\newblock \bibinfo{journal}{\emph{International Journal of Cognitive
  Ergonomics}} \bibinfo{volume}{4}, \bibinfo{number}{1} (\bibinfo{date}{March}
  \bibinfo{year}{2000}), \bibinfo{pages}{53--71}.
\newblock
\showISSN{1088-6362}
\urldef\tempurl%
\url{https://doi.org/10.1207/S15327566IJCE0401_04}
\showDOI{\tempurl}


\bibitem[\protect\citeauthoryear{Jordan, Shedden-Mora, and Löwe}{Jordan
  et~al\mbox{.}}{2017}]%
        {jordan_psychometric_2017}
\bibfield{author}{\bibinfo{person}{Pascal Jordan}, \bibinfo{person}{Meike~C.
  Shedden-Mora}, {and} \bibinfo{person}{Bernd Löwe}.}
  \bibinfo{year}{2017}\natexlab{}.
\newblock \showarticletitle{Psychometric analysis of the {Generalized}
  {Anxiety} {Disorder} scale ({GAD}-7) in primary care using modern item
  response theory}.
\newblock \bibinfo{journal}{\emph{PLoS ONE}} \bibinfo{volume}{12},
  \bibinfo{number}{8} (\bibinfo{date}{Aug.} \bibinfo{year}{2017}).
\newblock
\showISSN{1932-6203}
\urldef\tempurl%
\url{https://doi.org/10.1371/journal.pone.0182162}
\showDOI{\tempurl}


\bibitem[\protect\citeauthoryear{Kamita, Ito, Matsumoto, Munakata, and
  Inoue}{Kamita et~al\mbox{.}}{2019}]%
        {kamita_chatbot_2018}
\bibfield{author}{\bibinfo{person}{Takeshi Kamita}, \bibinfo{person}{Tatsuya
  Ito}, \bibinfo{person}{Atsuko Matsumoto}, \bibinfo{person}{Tsunetsugu
  Munakata}, {and} \bibinfo{person}{Tomoo Inoue}.}
  \bibinfo{year}{2019}\natexlab{}.
\newblock \showarticletitle{A {Chatbot} {System} for {Mental} {Healthcare}
  {Based} on {SAT} {Counseling} {Method}}.
\newblock \bibinfo{journal}{\emph{Mobile Information Systems}}
  \bibinfo{volume}{2019} (\bibinfo{date}{March} \bibinfo{year}{2019}),
  \bibinfo{pages}{9517321}.
\newblock
\showISSN{1574-017X}
\urldef\tempurl%
\url{https://doi.org/10.1155/2019/9517321}
\showDOI{\tempurl}


\bibitem[\protect\citeauthoryear{Kocaballi, Berkovsky, Quiroz, Laranjo, Tong,
  Rezazadegan, Briatore, and Coiera}{Kocaballi et~al\mbox{.}}{2019}]%
        {Baki2019}
\bibfield{author}{\bibinfo{person}{Ahmet~Baki Kocaballi},
  \bibinfo{person}{Shlomo Berkovsky}, \bibinfo{person}{Juan~C Quiroz},
  \bibinfo{person}{Liliana Laranjo}, \bibinfo{person}{Huong~Ly Tong},
  \bibinfo{person}{Dana Rezazadegan}, \bibinfo{person}{Agustina Briatore},
  {and} \bibinfo{person}{Enrico Coiera}.} \bibinfo{year}{2019}\natexlab{}.
\newblock \showarticletitle{The Personalization of Conversational Agents in
  Health Care: Systematic Review}.
\newblock \bibinfo{journal}{\emph{J Med Internet Res}} \bibinfo{volume}{21},
  \bibinfo{number}{11} (\bibinfo{date}{7 Nov} \bibinfo{year}{2019}),
  \bibinfo{pages}{e15360}.
\newblock
\showISSN{1438-8871}
\urldef\tempurl%
\url{https://doi.org/10.2196/15360}
\showDOI{\tempurl}


\bibitem[\protect\citeauthoryear{Kroenke, Spitzer, and Williams}{Kroenke
  et~al\mbox{.}}{2001}]%
        {kroenke_phq-9_2001}
\bibfield{author}{\bibinfo{person}{K. Kroenke}, \bibinfo{person}{R.~L.
  Spitzer}, {and} \bibinfo{person}{J.~B. Williams}.}
  \bibinfo{year}{2001}\natexlab{}.
\newblock \showarticletitle{The {PHQ}-9: validity of a brief depression
  severity measure}.
\newblock \bibinfo{journal}{\emph{Journal of General Internal Medicine}}
  \bibinfo{volume}{16}, \bibinfo{number}{9} (\bibinfo{date}{Sept.}
  \bibinfo{year}{2001}), \bibinfo{pages}{606--613}.
\newblock
\showISSN{0884-8734}
\urldef\tempurl%
\url{https://doi.org/10.1046/j.1525-1497.2001.016009606.x}
\showDOI{\tempurl}


\bibitem[\protect\citeauthoryear{Laranjo, Dunn, Tong, Kocaballi, Chen, Bashir,
  Surian, Gallego, Magrabi, Lau, and Coiera}{Laranjo et~al\mbox{.}}{2018}]%
        {laranjo}
\bibfield{author}{\bibinfo{person}{Liliana Laranjo}, \bibinfo{person}{Adam~G
  Dunn}, \bibinfo{person}{Huong~Ly Tong}, \bibinfo{person}{Ahmet~Baki
  Kocaballi}, \bibinfo{person}{Jessica Chen}, \bibinfo{person}{Rabia Bashir},
  \bibinfo{person}{Didi Surian}, \bibinfo{person}{Blanca Gallego},
  \bibinfo{person}{Farah Magrabi}, \bibinfo{person}{Annie Y~S Lau}, {and}
  \bibinfo{person}{Enrico Coiera}.} \bibinfo{year}{2018}\natexlab{}.
\newblock \showarticletitle{{Conversational agents in healthcare: a systematic
  review}}.
\newblock \bibinfo{journal}{\emph{Journal of the American Medical Informatics
  Association}} \bibinfo{volume}{25}, \bibinfo{number}{9} (\bibinfo{date}{07}
  \bibinfo{year}{2018}), \bibinfo{pages}{1248--1258}.
\newblock
\showISSN{1527-974X}
\urldef\tempurl%
\url{https://doi.org/10.1093/jamia/ocy072}
\showDOI{\tempurl}
\showeprint{https://academic.oup.com/jamia/article-pdf/25/9/1248/25643433/ocy072.pdf}


\bibitem[\protect\citeauthoryear{Ly, Ly, and Andersson}{Ly
  et~al\mbox{.}}{2017}]%
        {kien_fully_2017}
\bibfield{author}{\bibinfo{person}{Kien~Hoa Ly}, \bibinfo{person}{Ann-Marie
  Ly}, {and} \bibinfo{person}{Gerhard Andersson}.}
  \bibinfo{year}{2017}\natexlab{}.
\newblock \showarticletitle{A fully automated conversational agent for
  promoting mental well-being: {A} pilot {RCT} using mixed methods}.
\newblock \bibinfo{journal}{\emph{Internet Interventions}}
  \bibinfo{volume}{10} (\bibinfo{date}{Dec.} \bibinfo{year}{2017}),
  \bibinfo{pages}{39--46}.
\newblock
\showISSN{2214-7829}
\urldef\tempurl%
\url{https://doi.org/10.1016/j.invent.2017.10.002}
\showDOI{\tempurl}


\bibitem[\protect\citeauthoryear{Maharjan, Bækgaard, and Bardram}{Maharjan
  et~al\mbox{.}}{2019}]%
        {maharjan_hear_2019}
\bibfield{author}{\bibinfo{person}{Raju Maharjan}, \bibinfo{person}{Per
  Bækgaard}, {and} \bibinfo{person}{Jakob~E. Bardram}.}
  \bibinfo{year}{2019}\natexlab{}.
\newblock \showarticletitle{"{Hear} me out": smart speaker based conversational
  agent to monitor symptoms in mental health}. In
  \bibinfo{booktitle}{\emph{Adjunct {Proceedings} of the 2019 {ACM}
  {International} {Joint} {Conference} on {Pervasive} and {Ubiquitous}
  {Computing} and {Proceedings} of the 2019 {ACM} {International} {Symposium}
  on {Wearable} {Computers}}} \emph{(\bibinfo{series}{{UbiComp}/{ISWC} '19
  {Adjunct}})}. \bibinfo{publisher}{Association for Computing Machinery},
  \bibinfo{address}{London, United Kingdom}, \bibinfo{pages}{929--933}.
\newblock
\showISBNx{978-1-4503-6869-8}
\urldef\tempurl%
\url{https://doi.org/10.1145/3341162.3346270}
\showDOI{\tempurl}


\bibitem[\protect\citeauthoryear{Morris, Schueller, and Picard}{Morris
  et~al\mbox{.}}{2015}]%
        {morris_efficacy_nodate}
\bibfield{author}{\bibinfo{person}{Robert~R. Morris},
  \bibinfo{person}{Stephen~M. Schueller}, {and} \bibinfo{person}{Rosalind~W.
  Picard}.} \bibinfo{year}{2015}\natexlab{}.
\newblock \showarticletitle{Efficacy of a {Web}-based, crowdsourced
  peer-to-peer cognitive reappraisal platform for depression: randomized
  controlled trial}.
\newblock \bibinfo{journal}{\emph{Journal of Medical Internet Research}}
  \bibinfo{volume}{17}, \bibinfo{number}{3} (\bibinfo{date}{March}
  \bibinfo{year}{2015}), \bibinfo{pages}{e72}.
\newblock
\showISSN{1438-8871}
\urldef\tempurl%
\url{https://doi.org/10.2196/jmir.4167}
\showDOI{\tempurl}


\bibitem[\protect\citeauthoryear{Ng, Firth, Minen, and Torous}{Ng
  et~al\mbox{.}}{[n.d.]}]%
        {ngUserEngagementMental2019}
\bibfield{author}{\bibinfo{person}{Michelle~M. Ng}, \bibinfo{person}{Joseph
  Firth}, \bibinfo{person}{Mia Minen}, {and} \bibinfo{person}{John Torous}.}
  \bibinfo{year}{[n.d.]}\natexlab{}.
\newblock \showarticletitle{User {{Engagement}} in {{Mental Health Apps}}: {{A
  Review}} of {{Measurement}}, {{Reporting}}, and {{Validity}}}.
\newblock  \bibinfo{volume}{70}, \bibinfo{number}{7}
  (\bibinfo{year}{[n.\,d.]}), \bibinfo{pages}{538--544}.
\newblock
\showISSN{1075-2730}
\urldef\tempurl%
\url{https://doi.org/10/gg39pj}
\showDOI{\tempurl}


\bibitem[\protect\citeauthoryear{Ravichander and Black}{Ravichander and
  Black}{[n.d.]}]%
        {ravichander_proceedings_2018}
\bibfield{author}{\bibinfo{person}{A Ravichander} {and} \bibinfo{person}{A
  Black}.} \bibinfo{year}{[n.d.]}\natexlab{}.
\newblock \bibinfo{booktitle}{\emph{Proceedings Of The 19Th Annual Sigdial
  Meeting On Discourse And Dialogue.}}
\newblock


\bibitem[\protect\citeauthoryear{Stiles-Shields}{Stiles-Shields}{[n.d.]}]%
        {stiles-shields_woebot_2019}
\bibfield{author}{\bibinfo{person}{Colleen Stiles-Shields}.}
  \bibinfo{year}{[n.d.]}\natexlab{}.
\newblock \showarticletitle{Woebot: A Professional Review}.
\newblock  (\bibinfo{year}{[n.\,d.]}).
\newblock
\urldef\tempurl%
\url{https://onemindpsyberguide.org/expert-review/woebot-an-expert-review/}
\showURL{%
\tempurl}


\bibitem[\protect\citeauthoryear{Vaidyam, Wisniewski, Halamka, Kashavan, and
  Torous}{Vaidyam et~al\mbox{.}}{2019}]%
        {vaidyam_chatbots_2019}
\bibfield{author}{\bibinfo{person}{Aditya~Nrusimha Vaidyam},
  \bibinfo{person}{Hannah Wisniewski}, \bibinfo{person}{John~David Halamka},
  \bibinfo{person}{Matcheri~S. Kashavan}, {and} \bibinfo{person}{John~Blake
  Torous}.} \bibinfo{year}{2019}\natexlab{}.
\newblock \showarticletitle{Chatbots and {Conversational} {Agents} in {Mental}
  {Health}: {A} {Review} of the {Psychiatric} {Landscape}}.
\newblock \bibinfo{journal}{\emph{Canadian Journal of Psychiatry. Revue
  Canadienne De Psychiatrie}} \bibinfo{volume}{64}, \bibinfo{number}{7}
  (\bibinfo{year}{2019}), \bibinfo{pages}{456--464}.
\newblock
\showISSN{1497-0015}
\urldef\tempurl%
\url{https://doi.org/10.1177/0706743719828977}
\showDOI{\tempurl}


\bibitem[\protect\citeauthoryear{Woodward, Kanjo, Brown, McGinnity, Inkster,
  Macintyre, and Tsanas}{Woodward et~al\mbox{.}}{2019}]%
        {woodward_beyond_2019}
\bibfield{author}{\bibinfo{person}{K. Woodward}, \bibinfo{person}{E. Kanjo},
  \bibinfo{person}{D. Brown}, \bibinfo{person}{T.~M. McGinnity},
  \bibinfo{person}{B. Inkster}, \bibinfo{person}{D.~J. Macintyre}, {and}
  \bibinfo{person}{A. Tsanas}.} \bibinfo{year}{2019}\natexlab{}.
\newblock \showarticletitle{Beyond mobile apps: a survey of technologies for
  mental well-being}.
\newblock \bibinfo{journal}{\emph{IEEE Transactions on Affective Computing}}
  (\bibinfo{year}{2019}).
\newblock
\showISSN{1949-3045}
\urldef\tempurl%
\url{http://irep.ntu.ac.uk/id/eprint/36543/}
\showURL{%
\tempurl}


\end{thebibliography}

\end{document}